\documentstyle[aps,preprint,floats,epsf,epsfig]{revtex}
\begin{document}
\def\be{\begin{eqnarray}}
\def\en{\end{eqnarray}}
\def\non{\nonumber}
\def\la{\langle}
\def\ra{\rangle}
\def\nc{N_c^{\rm eff}}
\def\vp{\varepsilon}
\def\B{{\cal B}}
\def\up{\uparrow}
\def\dw{\downarrow}
\def\vma{{_{V-A}}}
\def\vpa{{_{V+A}}}
\def\smp{{_{S-P}}}
\def\spp{{_{S+P}}}
\def\J{{J/\psi}}
\def\ov{\overline}
\def\Lqcd{{\Lambda_{\rm QCD}}}
\def\pr{{\sl Phys. Rev.}~}
\def\prl{{\sl Phys. Rev. Lett.}~}
\def\pl{{\sl Phys. Lett.}~}
\def\np{{\sl Nucl. Phys.}~}
\def\zp{{\sl Z. Phys.}~}
\def\lsim{ {\ \lower-1.2pt\vbox{\hbox{\rlap{$<$}\lower5pt\vbox{\hbox{$\sim$}
}}}\ } }
\def\gsim{ {\ \lower-1.2pt\vbox{\hbox{\rlap{$>$}\lower5pt\vbox{\hbox{$\sim$}
}}}\ } }

\font\el=cmbx10 scaled \magstep2{\obeylines\hfill August, 2002}

\vskip 1.5 cm

\centerline{\large\bf Three-Body Charmful Baryonic $B$ Decays $\ov
B\to D(D^*)N\ov N$}
\bigskip
\centerline{\bf Hai-Yang Cheng$^{1}$ and Kwei-Chou Yang$^{2}$}
\medskip
\centerline{$^1$ Institute of Physics, Academia Sinica}
\centerline{Taipei, Taiwan 115, Republic of China}
\medskip

\medskip
\centerline{$^2$ Department of Physics, Chung Yuan Christian
University} \centerline{Chung-Li, Taiwan 320, Republic of China}
\bigskip
\bigskip
\centerline{\bf Abstract}
\bigskip
{\small We study the charmful three-body baryonic $B$ decays $\ov
B\to D^{(*)}N\ov N$: the color-allowed modes $\ov B^0\to
D^{(*)+}n\bar p$ and the color-suppressed ones $\ov B^0\to
D^{(*)0}p\bar p$. While the $D^{*+}/D^+$ production ratio is
predicted to be of order 3, it is found that $D^0p\bar p$ has a
similar rate as $D^{*0}p\bar p$. It is pointed out that $\ov
B^0\to D(D^*)N\ov N$ are dominated by the nucleon vector current
or by vector meson intermediate states, whereas $\ov B^0\to
D^0p\bar p$ proceeds mainly via the exchange of the axial-vector
intermediate state $a_1(1260)$. The study of the $N\bar N$
invariant mass distribution in general indicates a threshold
baryon pair production; that is, a recoil charmed meson
accompanied by a low mass baryon pair except that the spectrum of
$D^0p\bar p$ has a hump at large $p\bar p$ invariant mass
$m_{p\bar p}\sim 3.0$ GeV.

}

\pagebreak

\section{Introduction}
Previously CLEO has reported the first observation of
color-allowed charmful baryonic $B$ decays with sizable rates
\cite{CLEO}:
 \be
 \B(\ov B^0\to D^{*+}n\bar p) &=& (14.5^{+3.4}_{-3.0}\pm 2.7)\times
 10^{-4}, \non \\
 \B(\ov B^0\to D^{*+}p\bar p\pi^-) &=& (6.5^{+1.3}_{-1.2}\pm
 1.0)\times 10^{-4}.
 \en
This, when combining with the non-observation of the two-body
baryonic $B$ decays such as $B\to N\ov N$ \cite{CLEOBelleBaryon},
implies the dominance of multi-body final states in decays of $B$
mesons into baryons. Recently, Belle announced a similar
measurement for color-suppressed baryonic decay decays at the
level of $10^{-4}$ \cite{Belle}:
 \be
 \B(\ov B^0\to D^{*0}p\bar p) &=& (1.20^{+0.33}_{-0.29}\pm
 0.21)\times 10^{-4}, \non \\
 \B(\ov B^0\to D^{0}p\bar p) &=& (1.18\pm0.15\pm0.16)\times
 10^{-4},
 \en
with $5.6\sigma$ and $12\sigma$ statistical significance
respectively. Roughly speaking, the $D^{*0}p\bar p$ rate is
smaller than that of $D^{*+}n\bar p$ by one order of magnitude.

Another class of charmful baryonic $B$ decays is $\ov
B\to\Lambda_c(\Sigma_c)\ov N X$. The early CLEO measurement
\cite{CLEOc} and the new Belle \cite{Bellebaryon1} and CLEO
\cite{CLEObaryon} results show that the three-body charmful decay
$B^-\to\Lambda_c\bar p\pi^-$ has a magnitude larger than $\ov
B^0\to\Lambda_c\bar p$. These modes have been theoretically
studied in \cite{CYcharmful}. The recent first observation of the
penguin-dominated charmless baryonic decay $B^-\to p\bar p K^-$ by
Belle \cite{Bellebaryon} clearly indicates that it has a much
larger rate than the two-body counterpart $\ov B^0\to p\bar p$.
Theoretically, it has been explained in \cite{CYBaryon} why some
charmless three-body final states in which baryon-antibaryon pair
production is accompanied by a meson have a rate larger than their
two-body counterparts.

Under the factorization assumption, the decay amplitude, dominated
by the color-allowed external $W$-emission, is proportional to
$a_1\la O_1\ra_{\rm fact}$ where $O_1$ is a charged
current--charged current 4-quark operator, while the decay
amplitude, governed by the factorizable color-suppressed internal
$W$-emission, is described by $a_2\la O_2\ra_{\rm fact}$ with
$O_2$ being a neutral current--neutral current 4-quark operator.
Since $\ov B^0\to D^{(*)+}n\bar p$ are color-allowed, while $\ov
B^0\to D^{(*)+}n\bar p$ are color-suppressed, it is naively
expected that the measured ratio $R=\Gamma(\ov B^0\to
D^{(*)0}p\bar p)/\Gamma(\ov B^0\to D^{(*)+}n\bar p)$ can be used
to extract the parameter $|a_2/a_1|$, just as in the case of $\ov
B^0\to D^0\pi^0$ and $\ov B^0\to D^+\pi^-$ decays. However, there
is one complication here: the factorizable decay amplitude of
color-suppressed baryonic $B$ decays $\ov B^0\to D^{(*)0}p\bar p$
involves a three-body hadronic matrix element which is basically
unknown. Therefore, one needs to impose further theoretical
assumptions in order to extract $a_2$ from the color-suppressed
baryonic $B$ decay modes. The color-favored decays $\ov B^0\to
D^{(*)+}n\bar p$ have been studied in \cite{CHT1}. It turns out
that $D^{*+}/D^+$ production ratio is predicted to be of order 3.
It is thus anticipated that the $D^{*0}/D^0$ production ratio in
color suppressed decays is also of order 3. However,
experimentally the latter is consistent with unitary. This motives
us to investigate why $D^0p\bar p$ has a similar rate as
$D^{*0}p\bar p$.

All $\ov B\to D^{(*)}N\ov N$ decays can be described in terms of
the pole model; they receive contributions from various
intermediate states: vector mesons such as $\rho,~\omega$,
axial-vector mesons $a_1(1260)$, $f_1(1285)$, $f_1(1420)$, and
pseudoscalar mesons $\pi,\eta,\eta'$. It appears that the decay
$\ov B^0\to D^0p\bar p$ is very special: it is dominated by the
axial-vector meson states, whereas the other modes proceed mainly
through the vector meson poles. This enables us to explain the
similar rates for $D^0p\bar p$ and $D^{*0}p\bar p$.

This paper is organized as follows. We first discuss the
color-favored modes $\ov B^0\to D^{(*)+}n\bar p$ in Sec. 2 and
then turn to the color-suppressed ones $\ov B^0\to D^{(*)0}p\bar
p$ in Sec. 3. Discussions and conclusions are presented in Sec. 4.

\section{Color-allowed $\ov B^0\to D^{(*)+}{\lowercase{n\bar
p}}$} At the quark level, the color-allowed decays $\ov B^0\to
D^{+(*)}n\bar p$ proceed through the factorizable external
$W$-emission and $W$-exchange diagrams, and the nonfactorizable
internal $W$-emission [see Fig. 1(a)], while $\ov B^0\to
D^{0(*)}p\bar p$ via the factorizable internal $W$-emission,
$W$-exchange diagrams and the nonfactorizable internal
$W$-emission [see Fig. 2(a)]. More precisely, their factorizable
amplitudes read
 \be \label{famp}
 A(\ov B^0\to D^{+(*)}n\bar p)_{\rm fact} &=&
 {G_F\over\sqrt{2}}V_{ud}^*V_{cb}\Big\{ a_1\la n\bar p|(\bar
 du)|0\ra\la D^{+(*)}|(\bar c b)|\ov B^0\ra \non \\ &+& a_2\la D^{+(*)}n\bar
 p|(\bar cu)|0\ra\la 0|(\bar db)|\ov B^0\ra\Big\}, \non \\
  A(\ov B^0\to D^{0(*)}p\bar p)_{\rm fact} &=&
 {G_F\over\sqrt{2}}V_{ud}^*V_{cb}\Big\{ a_2\la n\bar p|(\bar
 db)|\ov B^0\ra\la D^{0(*)}|(\bar c u)|0\ra \non \\ &+& a_2\la D^{0(*)}p\bar
 p|(\bar cu)|0\ra\la 0|(\bar db)|\ov B^0\ra\Big\},
 \en
where $(\bar q_1q_2)\equiv \bar q_1\gamma_\mu(1-\gamma_5)q_2$, and
$a_1$, $a_2$, which will be specified later, are some
renormalization scale and scheme independent parameters . The
second term in each decay amplitude of Eq. (\ref{famp})
corresponds to the $W$-exchange amplitude, which is not only color
but also helicity suppressed. Since the three-body matrix element
$\la n\bar p|(\bar db)|\ov B^0\ra$ is basically unknown, we will
evaluate the color-suppressed amplitude based on the pole model
approximation.

\begin{figure}[tb]
\hspace{2cm} \psfig{figure=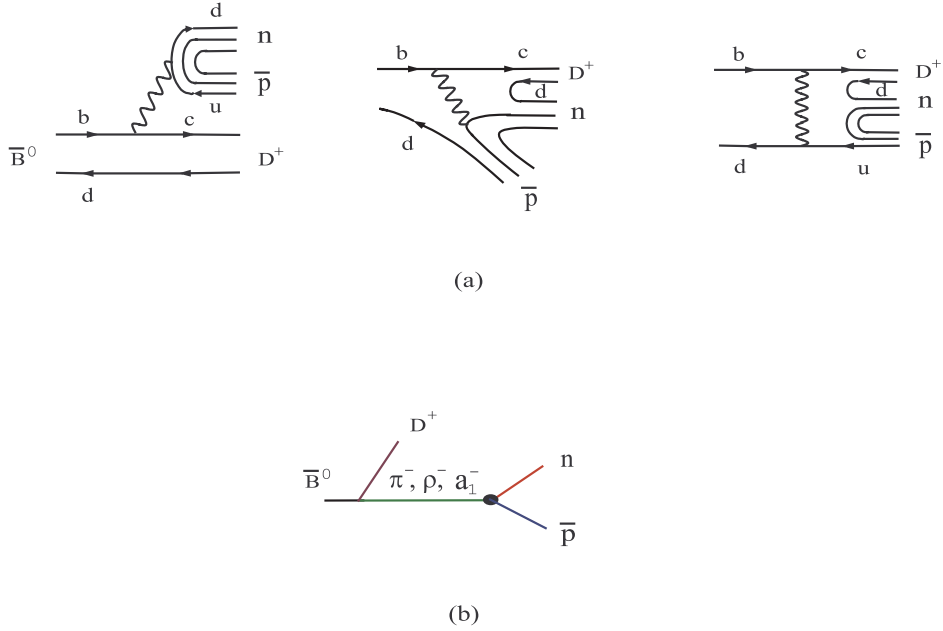,height=3.5in} \vspace{0.4cm}
    \caption{{\small (a) Quark diagrams for $\ov B^0\to D^+n\bar
    p$ and (b) the pole diagram for the factorizable external
    $W$-emission.
    }}
   \label{fig:1}
\end{figure}

\begin{figure}[tp]
\vspace{0cm} \hspace{2cm}\psfig{figure=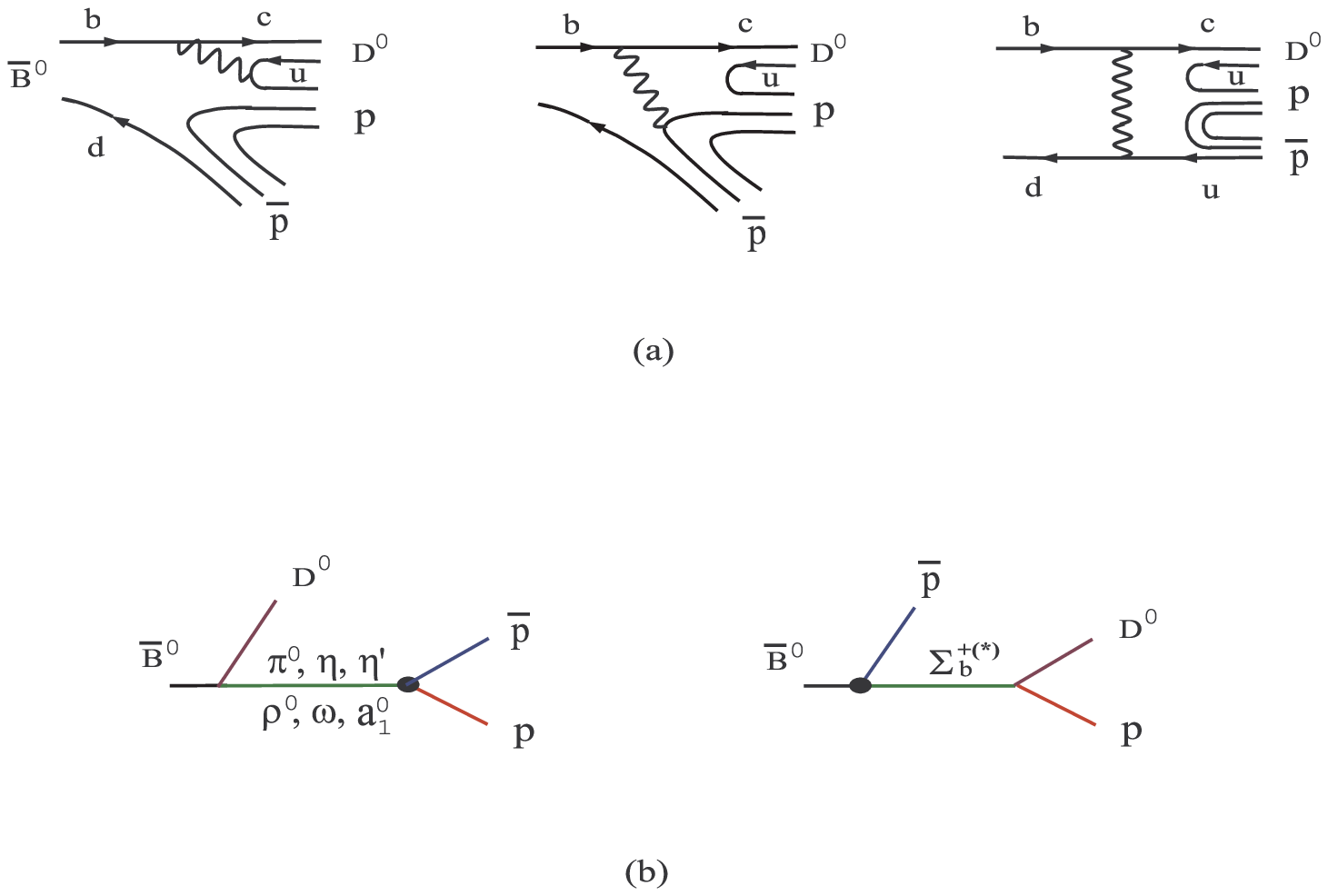,height=3.5in}
\vspace{0.5cm}
    \caption{{\small (a) Quark diagrams for $\ov B^0\to
    D^0p\bar p$ and (b) the pole diagrams for the factorizable
    internal $W$-emission.
    }}
   \label{fig:2}
\end{figure}

Let us first focus on the decays $\ov B^0\to D^{+(*)}n\bar p$. To
evaluate their factorizable amplitude, we need to know various
haronic matrix elements. The one-body and two-body mesonic matrix
elements are given by \cite{BSW}
 \be
\la P(q)|A_\mu|0\ra &=& -if_Pq_\mu, \qquad \la
V(p,\vp)|V_\mu|0\ra=f_Vm_V\vp^*_\mu,   \non \\
 \la P(p)|V_\mu|B(p_B)\ra &=& \left(p_{B\mu}+p_\mu-{m_B^2-m_{P}^2\over
q^2}\,q_ \mu\right) F^{BP}_1(q^2)+{m_B^2-m_{P}^2\over
q^2}q_\mu\,F^{BP}_0(q^2), \non
\\ \la V(p,\vp)|(V-A)_\mu|B(p_B)\ra &=& {2i\over
m_B+m_V}\,\epsilon_{\mu\nu\alpha \beta}\vp^{*\nu}p^\alpha
p_B^\beta V^{BV}(q^2)-\Bigg\{ (m_B+m_V)\vp^*_\mu A^{BV}_1(q^2)
\non
\\ &-& {\vp^*\cdot p_B\over m_B+m_V}\,(p_B+p)_\mu A^{BV}_2(q^2)
\non \\ &-& 2m_V\,{\vp^*\cdot p_B\over
q^2}\,q_\mu\big[A^{BV}_3(q^2)-A^{BV}_0(q^2)\big]\Bigg\},
 \en
where $q=p_B-p$, $F_1(0)=F_0(0)$, $A_3(0)=A_0(0)$, and
 \be
A_3(q^2)=\,{m_B+m_V\over 2m_V}\,A_1(q^2)-{m_B-m_V\over
2m_V}\,A_2(q^2).
 \en

The two-body baryonic matrix element appearing in Eq. (\ref{famp})
can be parametrized as
  \be \label{f,g}
\la n(p_n)\bar p(p_{\bar p})|(V- A)_\mu|0\ra &=& \bar
u_n(p_n)\Bigg\{f_1^{np}(q^2)\gamma_\mu+i{f_2^{np}(q^2)\over 2m_N}
\sigma_{\mu\nu}q^\nu+{f_3^{np}(q^2)\over 2m_N}q_\mu \non
\\ && -
\Big[g_1^{np}(q^2)\gamma_\mu+i{g_2^{np}(q^2)\over 2m_N}
\sigma_{\mu\nu}q^\nu+{g_3^{np}(q^2)\over
2m_N}q_\mu\Big]\gamma_5\Bigg\}v_{\bar p}(p_{\bar p}),
 \en
with $q=p_p+p_{\bar p}$. Among the six baryonic form factors, the
vector form factor $f_3(q^2)$ vanishes because of conservation of
vector current (CVC), while the absence of the second-class
current implies $g_2(q^2)=0$. Using CVC, the vector form factors
$f_{1,2}(q^2)$ can be related to the electromagnetic form factors
of the nucleon defined by
  \be \label{em}
 \la N(p_1)\ov N(p_2)|J_\mu^{\rm em}|0\ra=\bar
 u_N(p_1)\Big[F_1(q^2)\gamma_\mu+i{F_2(q^2)\over 2m_N}\sigma_{\mu\nu}q^\nu
 \Big]v_{\bar N}(p_2).
 \en
Specifically (see e.g. \cite{CYBaryon}),
 \be
 f^{np}_{1,2}(t)=F_{1,2}^p(t)-F_{1,2}^n(t),
 \en
where $t=q^2$.

The experimental data of the nucleon's e.m. form factors are
customarily described in terms of the electric and magnetic Sachs
form factors $G_E^N(t)$ and $G_M^N(t)$ which are related to
$F_1^N$ and $F_2^N$ via
 \be
 G_E^{p,n}(t)=F_1^{p,n}(t)+{t\over 4m_N^2}F_2^{p,n}(t), \qquad G_M^{p,n}(t)
 =F_1^{p,n}(t)+ F_2^{p,n}(t).
 \en
A recent phenomenological fit to the experimental data of nucleon
form factors has been carried out in \cite{CHT1} using the
following parametrization:
 \be \label{GMN}
 |G_M^p(t)| &=& \left({x_1\over t^2}+{x_2\over t^3}+{x_3\over t^4}
 +{x_4\over t^5}+{x_5\over t^6}\right)\left[\ln{t\over
 Q_0^2}\right]^{-\gamma},  \non \\
|G_M^n(t)| &=& \left({y_1\over t^2}+{y_2\over
t^3}\right)\left[\ln{t\over Q_0^2}\right]^{-\gamma},
 \en
where $Q_0=\Lambda_{\rm QCD}\approx$ 300 MeV and $\gamma=2+{4\over
3\beta}=2.148$\,. We will follow \cite{CHT2} to use the best fit
values
 \be
&& x_1=420.96\,{\rm GeV}^4, \qquad x_2=-10485.50\,{\rm GeV}^6,
 \qquad x_3=106390.97\,{\rm GeV}^8, \non \\
&& x_4=-433916.61\,{\rm GeV}^{10}, \qquad x_5=613780.15\,{\rm
GeV}^{12},
 \en
and
 \be \label{GMn1}
 y_1=236.69\,{\rm GeV}^4, \qquad\quad y_2=-579.51\,{\rm GeV}^6,
 \en
extracted from the neutron data under the assumption
$|G_E^n|=|G_M^n|$. Note that the form factors given by Eq.
(\ref{GMN}) do satisfy the constraint from perturbative QCD in the
limit of large $t$ \cite{CHT1}. Also as stressed in \cite{CHT1},
time-like magnetic form factors are expected to behave like
space-like magnetic form factors, i.e. real and positive for the
proton, but negative for the neutron.

In terms of the nucleon magnetic and electric form factors, the
weak form factors read
 \be \label{f12np}
 f^{np}_1(t) &=& {{t\over 4m_N^2}G_M^p(t)-G_E^p(t)\over t/(4m_N^2)-1}-
 {{t\over 4m_N^2}G_M^n(t)-G_E^n(t)\over t/(4m_N^2)-1}, \non \\
 f^{np}_2(t) &=& -{G_M^p(t)-G_E^p(t)\over t/(4m_N^2)-1}+
 {G_M^n(t)-G_E^n(t)\over t/(4m_N^2)-1}.
 \en
According to perturbative QCD, the weak form factors in the large
$t$ limit have the asymptotic expressions \cite{Brodsky}
 \be \label{larget}
 f^{np}_1(t) \to G_M^p(t)-G_M^n(t), \qquad
 g^{np}_1(t) \to {5\over 3}G_M^p(t)+G_M^n(t).
 \en
It is easily seen that this is consistent with the large $t$
behavior of $f_1^{np}$ given by Eq. (\ref{f12np}).

For the axial form factor $g_1(t)$, we shall follow \cite{CHT3} to
assume that it has a similar expression as $G_M^n(t)$
 \be \label{g1}
 g_1^{np}(t)=\left({d_1\over t^2}+{d_2\over t^3}\right)
 \left[\ln{t\over Q_0^2}\right]^{-\gamma},
 \en
where the coefficient $d_1$ is related to $x_1$ and $y_1$ by
considering  the asymptotic behavior of Sachs form factors $G_M^p$
and $G_M^n$ [see Eq. (\ref{GMN})]
 \be
 d_1={5\over 3}x_1-y_1.
 \en
As shown in \cite{CYBaryon}, the induced pseudoscalar form factor
$g_3$ corresponds to a pion pole contribution to the $n\bar p$
axial matrix element and it has the form
 \be
 g_3^{n p}(t)=-{4m_N^2\over t-m_\pi^2}g_1^{n p}(t).
 \en

\section{Color-suppressed $\ov B^0\to D^{(*)0}{\lowercase{p\bar
p}}$}

 We next turn to the color-suppressed modes $\ov B^0\to
D^{(*)0}p\bar p$ and assume that the main contributions arise from
the factorizable internal $W$-emission diagram [see Fig. 2(a)].
There are two corresponding pole diagrams as depicted in Fig.
2(b): one with bottom baryon poles $\Sigma_b({1\over 2}^+)$ and
$\Sigma_b^*({1\over 2}^-)$, and the other with meson poles. The
low-lying meson intermediate states are: $\pi^0,\eta,\eta'$,
$\rho^0,\omega$, and $a_1^0$ and possibly $f_1(1285)$ and
$f_1(1420)$.

\subsection{Meson pole contributions}
The meson pole contribution from Fig. 2 (b) is
 \be
 A(\ov B^0\to D^0p\bar p)_{\cal M} &=&
 {G_F\over\sqrt{2}}V_{ud}^*V_{cb}\,a_2\la D^0|(\bar
 cu)|0\ra\Bigg\{- \Big[ \la \pi^0|(\bar db)|\ov B^0\ra{i\over
 q^2-m_\pi^2}g_{\pi NN} \non \\
 &+& \la \eta|(\bar db)|\ov B^0\ra{i\over
 q^2-m_\eta^2}g_{\eta NN}+\la \eta'|(\bar db)|\ov B^0\ra{i\over
 q^2-m_{\eta'}^2}g_{\eta' NN}\Big]\bar u_p\gamma_5 v_{\bar p} \non \\
 &+& \la \rho^0|(\bar db)|\ov B^0\ra
 {i\over q^2-m_\rho^2}\,\bar
 u_p\,\vp_\rho^{\nu}\left(g_1^{\rho NN}\gamma_\nu+i{g_2^{\rho NN}\over
 2m_N}\sigma_{\nu\lambda}q^\lambda\right)v_{\bar p}  \non \\
  &+& \la \omega|(\bar db)|\ov B^0\ra
 {i\over q^2-m_\omega^2}\,\bar
 u_p\,\vp_\omega^{\nu}\left(g_1^{\omega NN}\gamma_\nu+i{g_2^{\omega NN}\over
 2m_N}\sigma_{\nu\lambda}q^\lambda\right)v_{\bar p}  \non \\
 &+& \la a^0_1|(\bar db)|\ov B^0\ra
 {i\over q^2-m_{a_1}^2}\,g_1^{a_1NN}\bar
 u_p\,\vp\!\!\!/_{a_1} \gamma_5 v_{\bar p} \Bigg\},
 \en
where $q=p_B-p_D=p_p+p_{\bar p}$. For simplicity, we have
concentrated on the low-lying poles and neglected those
contributions from the higher axial vector meson states such as
$f_1(1285)$ and $f_1(1420)$. As we shall see, the vector and
tensor coupling constants $g_1^{\rho NN}$ and $g_2^{\rho NN}$ are
related to the vector form factors $f_1^{np}$ and $f_2^{np}$
respectively, while $g_1^{a_1NN}$ and $g_{\pi NN}$ are connected
to the axial-vector form factors $g_1^{np}$ and $g_3^{np}$
respectively.

After some manipulation we obtain
 \be
  A(\ov B^0\to D^0p\bar p)_{\cal M} &=&
 -{G_F\over\sqrt{2}}V_{ud}^*V_{cb}\,f_D a_2\Bigg\{
 \Big[{1\over\sqrt{2}}
 (m_B^2-m_\pi^2)F_0^{B\pi}(m_D^2){g_{\pi NN}\over q^2-m_\pi^2}
 \non \\
 &+& (m_B^2-m_\eta^2)F_0^{B\eta}(m_D^2){g_{\eta NN}\over
 q^2-m_\eta^2}+  (m_B^2-m_{\eta'}^2)F_0^{B\eta'}(m_D^2){g_{\eta' NN}\over
 q^2-m_{\eta'}^2}\Big]\bar u_p\gamma_5 v_{\bar p}   \non \\
 &+& {\sqrt{2} m_\rho\over q^2-m_\rho^2} A_0^{B\rho}(m_D^2)\bar
 u_p\Big[-(g_1^{\rho NN}+g_2^{\rho NN})p\!\!\!/_B+{g_2^{\rho NN}\over
 2m_N}(p_p-p_{\bar p})\cdot p_B\Big]v_{\bar p} \non \\
 &+& {2 m_\omega\over q^2-m_\omega^2} A_0^{B\omega}(m_D^2)\bar
 u_p\Big[-(g_1^{\omega NN}+g_2^{\omega NN})p\!\!\!/_B+{g_2^{\omega NN}\over
 2m_N}(p_p-p_{\bar p})\cdot p_B\Big]v_{\bar p} \non \\
 &+& {\sqrt{2} m_{a_1}\over q^2-m_{a_1}^2} V_0^{Ba_1}(m_D^2)g_1^{a_1 NN}\bar
 u_p\Big[- p\!\!\!/_B+{p_B\cdot q\,q\!\!\!/\over m_{a_1}^2}\Big]\gamma_5v_{\bar  p}\Bigg\},
 \en
where we have applied the relations $\bar u_p q\!\!\!/ v_{\bar
p}=0$, $(p_p-p_{\bar p})\cdot q=0$ and employed the form factors
defined by
  \be \label{formBa1}
 \la a_1^-(p_{a_1})|(\bar ub)_\vma|\ov B^0(p_B)\ra &=& {2i\over
 m_B+m_{a_1}}\epsilon_{\mu\nu\alpha\beta}
\vp^{*\nu}p^\alpha_{a_1} p^\beta_B A^{Ba_1}(q^2)  \non \\ &-&
\Bigg\{(m_B+m_{a_1}) \vp^*_\mu V_1^{Ba_1}(q^2)  - {\vp^*\cdot
p_B\over m_B+m_{a_1}}(p_B+p_{a_1})_\mu V_2^{B\rho}(q^2) \non \\
&-& 2m_{a_1} {\vp^*\cdot p_B\over
q^2}q_\mu\left[V_3^{Ba_1}(q^2)-V_0^{Ba_1}(q^2)\right]\Bigg\},
 \en
with
  \be
V_3(q^2)=\,{m_B+m_V\over 2m_V}\,V_1(q^2)-{m_B-m_V\over
2m_V}\,V_2(q^2),
 \en
and $V_3(0)=V_0(0)$. Note that the pion in the form factor
$F_0^{B\pi}$ is referred to the charged one, so are the form
factors $A_0^{B\rho}$ and $V_0^{Ba_1}$.

Likewise, the meson pole contribution to $\ov B^0\to D^{*0}p\bar
p$ reads
  \be
  A(\ov B^0\to D^{*0}p\bar p)_{\cal M} &=&
 {G_F\over\sqrt{2}}V_{ud}^*V_{cb}\,a_2\,f_{D^*}m_{D^*}\Bigg\{ \Big[
 \sqrt{2}F_1^{B\pi}(m_{D^*}^2){g_{\pi NN}\over q^2-m_\pi^2}
 \non \\
 &+& 2F_1^{B\eta}(m_{D^*}^2){g_{\eta NN}\over
 q^2-m_\eta^2}+  2F_1^{B\eta'}(m_{D^*}^2){g_{\eta' NN}\over
 q^2-m_{\eta'}^2}\Big](\vp^*_{D^*}\cdot p_B)\bar u_p\gamma_5 v_{\bar p}   \non \\
 &+& \la \rho^0|\bar d\,\vp\!\!\!/_{D^*}^*(1-\gamma_5)b|\ov B^0\ra {1\over
 q^2-m_\rho^2}\bar u_p\,\vp_\rho^{\mu}(g_1^{\rho NN}\gamma_\mu+i{g_2^{\rho NN}\over
 2m_N}\sigma_{\mu\nu}q^\nu)v_{\bar p}  \non \\
  &+& \la \omega|\bar d\,\vp\!\!\!/_{D^*}^*(1-\gamma_5)b|\ov B^0\ra {1\over
 q^2-m_\omega^2}\bar u_p\,\vp_\omega^{\mu}(g_1^{\omega NN}\gamma_\mu+i{g_2^{\omega NN}\over
 2m_N}\sigma_{\mu\nu}q^\nu)v_{\bar p}  \non \\
  &+& \la a_1^0|\bar d\,\vp\!\!\!/_{D^*}^*(1-\gamma_5)b|\ov B^0\ra {1\over
 q^2-m_{a_1}^2}\bar u_p(g_1^{a_1 NN}\vp\!\!\!/_{a_1}\gamma_5)v_{\bar p}
 \Bigg\}.
 \en

\subsection{Baryon pole contributions}
In addition to the aforementioned meson pole contributions, there
also exist baryon pole diagrams, namely, the strong process $\ov
B^0\to \Sigma_b^{+(*)}\bar p$ followed by the weak decay
$\Sigma^{+(*)}_b\to D^0p$. Due to the large theoretical
uncertainties with the ${1\over 2}^-$ state $\Sigma_b^{+*}$, we
will focus on the ${1\over 2}^+$ intermediate state and its
amplitude is given by
 \be
 A(\ov B^0\to D^{0}p\bar p)_{\cal B}
 &=&{G_F\over\sqrt{2}}V_{ud}^*V_{cb} f_D\,a_2\,g_{\Sigma_b^+\to B\bar
 p}\,{1\over (p_p+p_D)^2-m_{\Sigma_b}^2 } \non \\
 &\times&
 \bar u_p\Big\{f_1^{\Sigma_b^+ p}(m_D^2)[2p_D\cdot
 p_p+p\!\!\!/_D(m_{\Sigma_b}-m_p)]\gamma_5  \non \\
 &+& g_1^{\Sigma_b^+ p}(m_D^2)[2p_D\cdot
 p_p-p\!\!\!/_D(m_{\Sigma_b}+m_p)]\Big\}
 v_{\bar p},
 \en
where we have applied the factorization approximation to the weak
decay $\Sigma_b^+\to D^0p$. Similarly, for $\ov B^0\to D^{*0}p\bar
p$ we have
  \be
 A(\ov B^0\to D^{*0}p\bar p)_{\cal B}
 &=&{G_F\over\sqrt{2}}V_{cb}V_{ud}^* f_{D^*}m_{D^*}\,a_2\,g_{\Sigma_b^+\to B\bar
 p}\,{1\over (p_p+p_{D^*})^2-m_{\Sigma_b}^2 } \non \\
 &\times& \bar u_p\,\vp^{*\mu}\Big\{f_1^{\Sigma_b^+ p}(m_{D^*}^ 2)[2p_{p\mu}+
 (m_{\Sigma_b}-m_p)\gamma_\mu+\gamma_\mu p\!\!\!/_D]\gamma_5  \non \\
 &+& g_1^{\Sigma_b^+ p}(m_{D^*}^2)[2p_{p\mu}-
 (m_{\Sigma_b}+m_p)\gamma_\mu+\gamma_\mu p\!\!\!/_D]\Big\}
 v_{\bar p},
 \en
where $\vp_\mu^*$ is the polarization vector of the $D^*$.

The baryon pole contribution is expected to be suppressed relative
to the meson pole due to the smallness of the strong coupling of
$\Sigma_b^+\to B^0\bar p$ \cite{CYBaryon}.

\section{Calculations and Results}
To proceed numerical calculations we first need to know the
relevant form factors, decay constants, strong couplings, and the
parameters $a_1$, $a_2$, which will be discussed in more detail
below.

\subsection{form factors and decay constants}
For the mesonic form factors of $B\to P$ and $B\to V$ transitions
we use the Melikhov-Stech (MS) model based on the constituent
quark picture \cite{MS} and the Neubert-Rieckert-Stech-Xu (NRSX)
model \cite{NRSX} which takes the Bauer-Stech-Wirbel (BSW) model
\cite{BSW} results for the form factors at zero momentum transfer
but makes a different ansatz for their $q^2$ dependence, namely, a
dipole behavior is assumed for the form factors
$F_1,~A_0,~A_2,~V$, motivated by heavy quark symmetry, and a
monopole dependence for $F_0,A_1$.

For $B\to a_1$ form factors, there are two existing calculations:
one in a quark-meson model \cite{Deandrea} and the other based on
the QCD sum rule \cite{Aliev}. The results are quite different,
for example, $V^{Ba_1}(0)$ computed in the quark-meson model,
1.20\,, is larger than the sum-rule prediction, $-0.23\pm 0.05$\,,
by a factor of five. We shall see later that $\ov B^0\to D^0p\bar
p$ is rather sensitive to the form factor $V^{Ba_1}_0$. It turns
that in order to accommodate the measurement of this decay, this
form factor should be around 0.86 which is between the
above-mentioned model calculations. In the present paper we shall
use the quark-meson model results for the $B\to a_1$ form factors
except that the value of $V^{Ba_1}_0(0)$ is replaced by 0.85
rather than 1.20\,.

To compute the form factors for $F_0^{B\eta}$ and $F_0^{B\eta'}$,
it is more natural to consider the flavor basis of $\eta_q$ and
$\eta_s$ defined by
 \be
 \eta_q={1\over\sqrt{2}}(u\bar u+d\bar d),\qquad\quad
 \eta_s=s\bar s.
 \en
The wave functions of the $\eta$ and $\eta'$ are given by
 \be \label{etawf}
 \left(\matrix{ \eta \cr \eta'\cr}\right)=\left(\matrix{ \cos\phi & -\sin\phi \cr
 \sin\phi & \cos\phi\cr}\right)\left(\matrix{\eta_q \cr \eta_s
 \cr}\right),
 \en
where $\phi=\theta+{\rm arctan}\sqrt{2}$, and $\theta$ is the
$\eta\!-\!\eta'$ mixing angle in the octet-singlet basis. The
physical form factors then have the simple expressions:
 \be \label{Beta}
F_{0,1}^{B\eta}={1\over\sqrt{2}}\cos\phi \,F_{0,1}^{B\eta_{u\bar
u}}, \qquad && F_{0,1}^{B\eta'}={1\over\sqrt{2}}\sin\phi\,
F_{0,1}^{B\eta'_{u\bar u}}.
 \en
Using $F_0^{B\eta_{u\bar u}}(0)=0.307$ and $F_0^{B\eta'_{u\bar
u}}(0)=0.254$ obtained from \cite{BSW} and the mixing angle
$\phi=39.3^\circ$ (or $\theta=-15.4^\circ$) \cite{Kroll} we find
$F_0^{B\eta}(0)=0.168$ and $F_0^{B\eta'}(0)=0.114$ in the BSW
model and hence the NRSX model. For other form-factor models, we
shall apply the relation based on the isospin-quartet symmetry
 \be \label{isospinquart}
 F_{0,1}^{B\eta_{u\bar u}}=F_{0,1}^{B\eta'_{u\bar
 u}}=F_{0,1}^{B\pi}
 \en
and Eq. (\ref{Beta}) to obtain the physical $B\to\eta$ and
$B\to\eta'$ transition form factors. For the MS model we obtain
$F_0^{B\eta}(0)=0.141$ and $F_0^{B\eta'}(0)=0.115$.

For the heavy-light baryonic form factors $f_1^{\Sigma_b^+p}$ and
$g_i^{\Sigma_b^+p}$, we will follow \cite{CT96} to apply the
nonrelativistic quark model to evaluate the weak current-induced
baryon form factors at zero recoil in the rest frame of the heavy
parent baryon, where the quark model is most trustworthy.
Following \cite{Cheng97} we have
 \be
 f_1^{\Sigma_b^+ p}(q^2_m)=1.703, \qquad\qquad
g_1^{\Sigma_b^+ p}(q^2_m)=-0.166
 \en
at zero recoil $q_m^2=(m_{\Sigma_b}-m_p)^2$. Since the calculation
for the $q^2$ dependence of form factors is beyond the scope of
the non-relativistic quark model, we will follow the conventional
practice to assume a pole dominance for the form-factor $q^2$
behavior:
 \be
 f(q^2)=f(q^2_m)\left({1-q^2_m/m^2_V\over 1-q^2/m_V^2} \right)^n\,,\qquad
 g(q^2)=g(q^2_m)
\left({1-q^2_m/m^2_A\over 1-q^2/m_A^2} \right)^n\,,
 \en
where $m_V$ ($m_A$) is the pole mass of the vector (axial-vector)
meson with the same quantum number as the current under
consideration.

For the decay constants we use $f_D=200$ MeV, $f_{D^*}=230$ MeV
and $f_{a_1}=205$ MeV.

\subsection{strong couplings}
In order to compute the decay rate for $\ov B^0\to D^{(*)0}p\bar
p$ we also need to know the strong couplings $g_{\pi
NN},\,g_1^{\rho NN},\,g_2^{\rho NN}$ and $g_1^{a_1NN}$ and their
$q^2$ dependence. To do this,  let us consider the pole
contributions to $\ov B^0\to D^{(*)+}n\bar p$. In the pole model
description, the relevant intermediate states are $\pi^-,~\rho^-$
and $a_1^-(1260)$ as shown in Fig. 1(b). The matrix element $\la
n\bar p|(V-A)_\mu|0\ra$ then reads
 \be
 \la n\bar p|(V-A)_\mu |0\ra_{\rm pole} &=& \bar u_n\Bigg\{
 {\sqrt{2}f_\rho m_\rho\over q^2-m_\rho^2}\left[g_1^{\rho
 NN}\gamma_\mu+i{g_2^{\rho NN}\over
 2m_N}\sigma_{\mu\nu}q^\nu\right] \non \\
 &-& {\sqrt{2}f_{a_1} m_{a_1}\over q^2-m_{a_1}^2}g_1^{a_1
 NN}\gamma_\mu\gamma_5 - {\sqrt{2}f_\pi g_{\pi NN}\over
 q^2-m_\pi^2} q_\mu\gamma_5\Bigg\} v_{\bar p}.
 \en
Comparing this with Eq. (\ref{f,g}) we see that the $\rho^-$ meson
is responsible for the vector form factors $f_1$ and $f_2$,
$a_1^-(1260)$ for $g_1$ and $g_2$, and $\pi^-$ for the induced
pseudoscalar form factor $g_3$. More precisely,
 \be
 g_1^{\rho NN}(q^2) &=& {q^2-m_\rho^2\over \sqrt{2}f_\rho
 m_\rho}\,f_1^{np}(q^2), \qquad\quad  g_2^{\rho NN}(q^2) ={q^2-m_\rho^2\over \sqrt{2}f_\rho
 m_\rho}\,f_2^{np}(q^2),  \non \\
 g_1^{a_1 NN}(q^2) &=& {q^2-m_{a_1}^2\over \sqrt{2}f_{a_1}
 m_{a_1}}\,g_1^{np}(q^2), \qquad\quad  g_{\pi NN}(q^2) = {(q^2-m_\pi^2)\over
 2\sqrt{2} f_\pi  m_N}\,g_3^{np}(q^2).
 \en
As for the vector and tensor couplings of the $\omega$ meson, {\it
a priori} they are not necessarily related to those of the $\rho$
meson. For simplicity we shall assume that $g_{1,2}^{\omega
NN}(q^2)=g_{1,2}^{\rho NN}(q^2)/\sqrt{2}$, noting that the $\rho$
meson here is referred to the charged one.

As for the strong $\eta$ and $\eta'$ couplings with nucleons, we
shall apply the $^3P_0$ quark-pair creation model
\cite{Yaouanc,Jarfi} to estimate its strength relative to the
pion. This model in which the $q\bar q$ pair is created from the
vacuum with vacuum quantum numbers $^3P_0$ implies
 \be
 {g_{\eta p\bar p}\over g_{\pi p\bar p}}={\la
 \Phi_{p^\up}(124)\Phi_{\eta}
 (35)|\Phi_{p^{\up}}(123)\Phi_{\rm vac}(45)\ra\over \la
\Phi_{p^\up}(124)\Phi_{\pi^0} (35)|\Phi_{p^{\up}}(123)\Phi_{\rm
vac}(45)\ra},
 \en
where the $\Phi$'s are the spin-flavor wave functions and the
vacuum wave function has the expression
 \be
 \Phi_{\rm vac}={1\over\sqrt{3}}(u\bar u+d\bar d+s\bar s)\otimes
 {1\over\sqrt{2}}(\up\dw+\dw\up).
 \en
Using the proton wave function
 \be
 p^{\up}&=& {1\over\sqrt{3}}[duu\chi_s+(12)+(13)],
 \en
with $abc\chi_s=(2a^\dw b^\up c^\up-a^\up b^\up c^\dw-a^\up b^\dw
c^\up)/\sqrt{6}$, the $\pi^0$ meson wave function
 \be
\Phi_{\pi^0}={1\over\sqrt{2}}(u\bar u-d\bar d)\otimes
{1\over\sqrt{2}}(\up\dw-\dw\up),
 \en
and the $\eta$ and $\eta'$ flavor wave functions given by Eq.
(\ref{etawf}), we obtain
 \be
 {g_{\eta NN}\over g_{\pi NN}}={3\over 5}\cos\phi, \qquad\quad
  {g_{\eta' NN}\over g_{\pi NN}}={3\over 5}\sin\phi.
 \en
Strictly speaking, the above relations hold only at low energies.
But we shall assume their validity at arbitrary $q^2$.

As for the strong coupling $g_{\Sigma_b^+\to\ov B^0p}$, we use the
experimental result for $B^-\to\Lambda_c\bar p\pi^-$ to fix the
absolute coupling strength of $g_{\Lambda_b^+\to\ov B^0p}$ which
is in turn related to $g_{\Sigma_b^+\to\ov B^0p}$ via the $^3P_0$
quark-pair-creation model \cite{CYcharmful}. It is found that
$|g_{\Sigma_b^+\to\ov B^0p}|\sim 5$.

\subsection{$a_1$ and $a_2$}
In the naive factorization approach, the parameters $a_1$ and
$a_2$ are given by $a_{1,2}=c_{2,1}+c_{1,2}/N_c$, but this does
not include nonfactorizable effects which are especially important
for $a_2$. Phenomenologically, one can treat $a_{1,2}$ as free
parameters and extract them from experiment. The experimental
measurement of $B\to\J K$ leads to $|a_2(\J K)|=0.26\pm 0.02$
\cite{a1a2}. This seems to be also supported by the study of $B\to
D\pi$ decays: Assuming no relative phase between $a_1$ and $a_2$,
the result $a_2\sim {\cal O}(0.20-0.30)$ \cite{a1a2,ns} is
inferred from the data of $\ov B^0\to D^{(*)+}\pi^-$ and $B^-\to
D^{(*)0}\pi^-$. However, the above value of $a_2$ leads to too
small decay rates for $\ov B^0\to D^{(*)0}\pi^0$ when compared to
recent measurements by Belle and CLEO \cite{BelleCLEO}. In order
to account for the observation, one needs a larger $a_2(D\pi)$
with a non-trivial phase relative to $a_1$
\cite{Xing,Chenga1a2,NP}.

Using the measurements of CLEO and Belle for $\ov B^0\to
D^{(*)0}\pi^0$ \cite{BelleCLEO}, the magnitudes of $a_1$ and $a_2$
and their relative phase are extracted in \cite{Chenga1a2}, as
exhibited in Table I. We will use the values of $a_{1,2}(D\pi)$ to
compute the decay rate for $\ov B^0\to D^+n\bar p, D^0p\bar p$ and
$a_{1,2}(D^*\pi)$ for $\ov B^0\to D^{*+}n\bar p, D^{*0}p\bar p$.

{\squeezetable
\begin{table}[ht]
\caption{Extraction of the parameters $a_1$ and $a_2$ from the
measured $B\to D^{(*)}\pi$ rates by assuming a negligible
$W$-exchange contribution. Note that $a_2(D\pi)$ and $a_2(D^*\pi)$
should be multiplied by a factor of (200 MeV/$f_D$) and (230
MeV/$f_{D^*}$), respectively. This table is taken from [28].
 }
\footnotesize
\begin{center}
\begin{tabular}{l | c c c | c c c }
 Model~~ & $|a_1(D\pi)|$ & $|a_2(D\pi)|$ & $a_2(D\pi)/a_1(D\pi)$~ & $|a_1(D^*\pi)|$
 & $|a_2(D^*\pi)|$ & $a_2(D^*\pi)/a_1(D^*\pi)$ \\ \hline
 NRSX & $0.85\pm0.06$ & $0.40\pm0.05$ & $(0.47\pm0.05)\,{\rm exp}(i59^\circ)$
 & $0.94\pm0.04$ & $0.31\pm0.04$ & $(0.33\pm0.04)\,{\rm exp}(i63^\circ)$ \\
 MS & $0.88\pm0.06$ & $0.47\pm0.06$ & $(0.53\pm0.06)\,{\rm exp}(i59^\circ)$
 & $0.85\pm0.03$ & $0.39\pm0.05$ & $(0.46\pm0.06)\,{\rm exp}(i63^\circ)$ \\
\end{tabular}
\end{center}
\end{table}
}

\subsection{Results}
In principle, the unknown parameter $d_2$ appearing in the form
factor $g_1^{np}(q^2)$ [cf. Eq. (\ref{g1})] can be fitted to the
measured central value of the branching ratio for $\ov B^0\to
D^{*+}n\bar p$ as it is theoretically much more clean. However, we
find that the decay rate of $\ov B^0\to D^0p\bar p$ is dominated
by the axial-vector meson poles and hence it is rather sensitive
to $g^{np}_1(t)$ and hence $d_2$. Therefore, we instead fix it by
fitting to the measured central value of $\B(\ov B^0\to D^0p\bar
p)=1.18\times 10^{-4}$. We obtain $d_2=-2070\,{\rm GeV}^6$ and
$-2370\,{\rm GeV}^6$, respectively, in NRSX and MS form-factor
models.

The total decay rate for the process $\ov B(p_B)\to N(p_1)+\ov
N(p_2)+D(p_3)$ is computed by the formula
 \be
 \Gamma = {1\over (2\pi)^3}\,{1\over 32m_B^3}\int |A|^2dm_{12}^2dm_{23}^2,
 \en
where $m_{ij}^2=(p_i+p_j)^2$ with $p_3=p_D$. To compute the
branching ratios, we use the $B$ meson lifetimes quoted in
\cite{PDG}.

\begin{table}[ht]
\caption{Branching ratios (in units of $10^{-4}$) for charmful
decays $\ov B^0\to D^{(*)+}n\bar p$ and $\ov B^0\to D^{(*)0}p\bar
p$ calculated in the MS and NRSX form-factor models. The first
(second) number in parentheses is the branching ratio due to the
vector (axial-vector and pseudoscalar) current or intermediate
vector (axial-vector) meson contributions.}
\begin{center}
\begin{tabular}{l c c l }
Decay & MS & NRSX & Expt. \cite{CLEO,Belle} \\ \hline
 $\ov B^0\to D^+ n\bar p$ & $3.17~(3.04,0.12)$ & $3.64~(3.47,0.16)$  &  \\
 $\ov B^0\to D^{*+}n\bar p$ & $10.0~(9.54,0.49)$ & $11.0~(10.2,0.76)$
 & $14.5^{+3.4}_{-3.0}\pm 2.7$ \\
 $\ov B^0\to D^0p\bar p$ & $1.18~(0.15,1.03)$ & $1.17~(0.11,1.06)$
 & $1.18\pm0.15\pm0.16$ \\
 $\ov B^0\to D^{*0}p\bar p$ & $1.58~(1.42,0.17)$ & $1.23~(1.12,0.11)$
 & $1.20^{+0.33}_{-0.29}\pm  0.21$  \\
\end{tabular}
\end{center}
\end{table}

The results are shown in Table II. As stated before, we fit the
unknown parameter $d_2$ to the measured branching ratio of $\ov
B^0\to D^0p\bar p$ and then in turn predict other neutral baryonic
$B$ modes. The baryon pole contributions to $D^{(*)0}p\bar p$ are
found to be at most of order of $10^{-6}$ and hence they are
negligible. It is clear from Table II that the predicted rates are
consistent with experiment. We see that $\ov B^0\to D(D^*)N\ov N$
are dominated by the vector current or by vector meson
intermediate states,\footnote{As far as the axial vector
contribution to the branching ratio is concerned, our result for
$\ov B^0\to D^{*+}n\bar p$ is quite different from that given in
\cite{CHT3}, though the value of $d_2$ is similar. For example,
$\B_A\sim 12.7\times 10^{-4}$ is obtained in \cite{CHT3}, whereas
it is only $0.6\times 10^{-4}$ in our case. If $g_1^{np}$ is
identified with the asymptotic form ${5\over 3}G_M^p+G_M^n$ in the
whole time-like region [cf. Eq.(\ref{larget})], $\B_A$ in our case
will be of order only $8\times 10^{-6}$, while it can be as large
as $1\times 10^{-4}$ in \cite{CHT3}.} whereas $\ov B^0\to D^0p\bar
p$ is dominated by the axial-vector intermediate state
$a_1(1260)$. Note that the ratio $D^{*+}/D^+$ is of order 3, while
$D^{*0}/D^0$ is close to unity.

In Fig. 3 we show the $n\bar p$ invariant mass distributions
$d\B/dm_{n\bar p}$ of $\ov B^0\to D^{*+}n\bar p$ and $\ov B^0\to
D^+n\bar p$, where $m_{n\bar p}$ is the invariant mass of the
nucleon pair. Evidently, the spectrum peaks at $m_{p\bar p}\sim
2.1\,{\rm GeV}$, indicating a threshold enhancement for baryon
production, that is, a  recoil charmed meson is accompanied by a
nucleon pair with low invariant mass. This effect is due to the
suppression of the baryonic form factors at large $t$. Physically,
this can be visualized that the quark and anti-quark forming a
nucleon pair are moving collinearly and energetically, so that the
invariant mass $m_{p\bar p}$ tends to be small and near threshold.

\begin{figure}[tb]
\begin{tabular}{cc}
\hspace{-8cm}\scalebox{1}{\epsfig{file=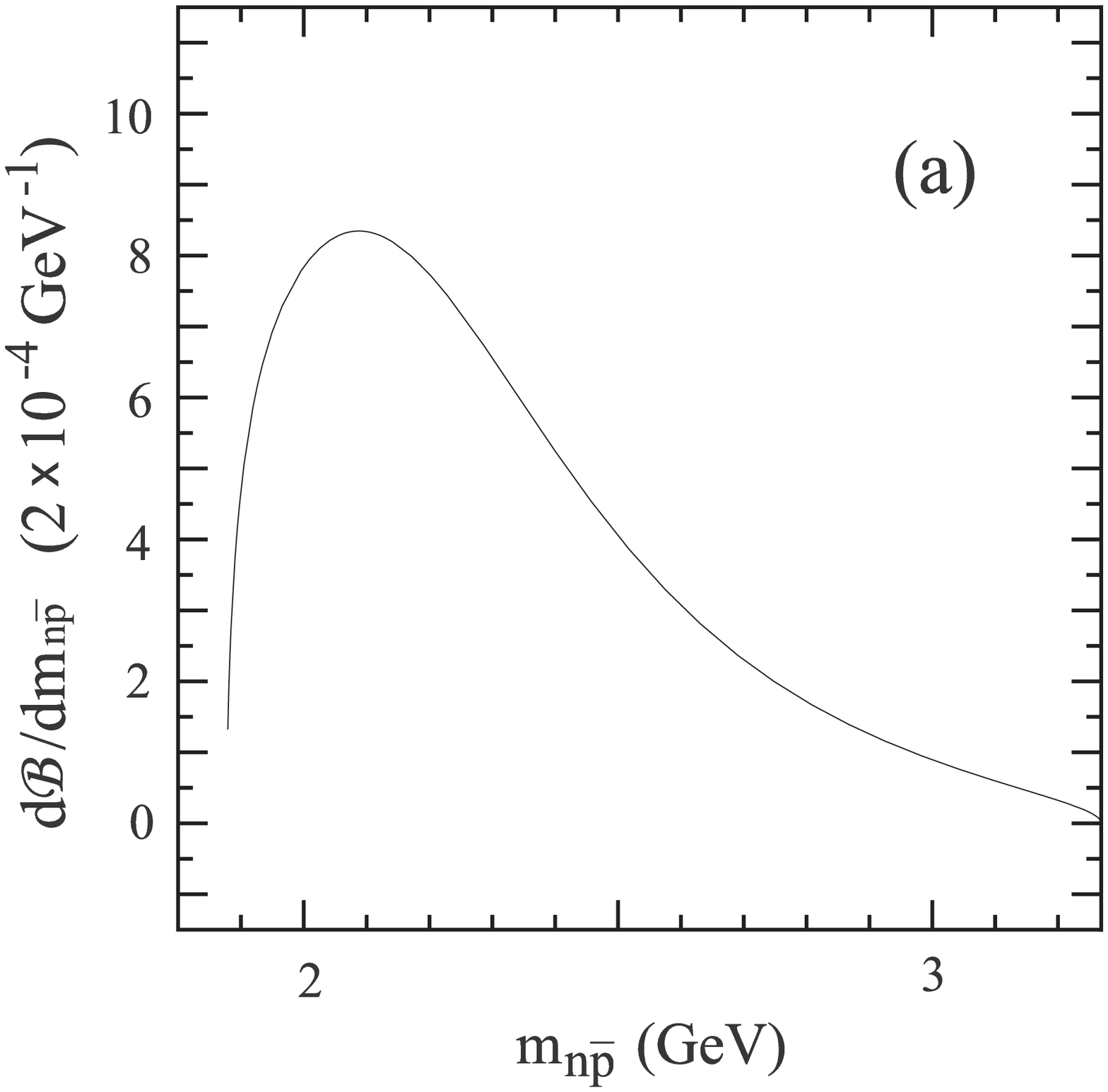,height=2.2in}}
& \hspace{-8cm}
\scalebox{1}{\epsfig{file=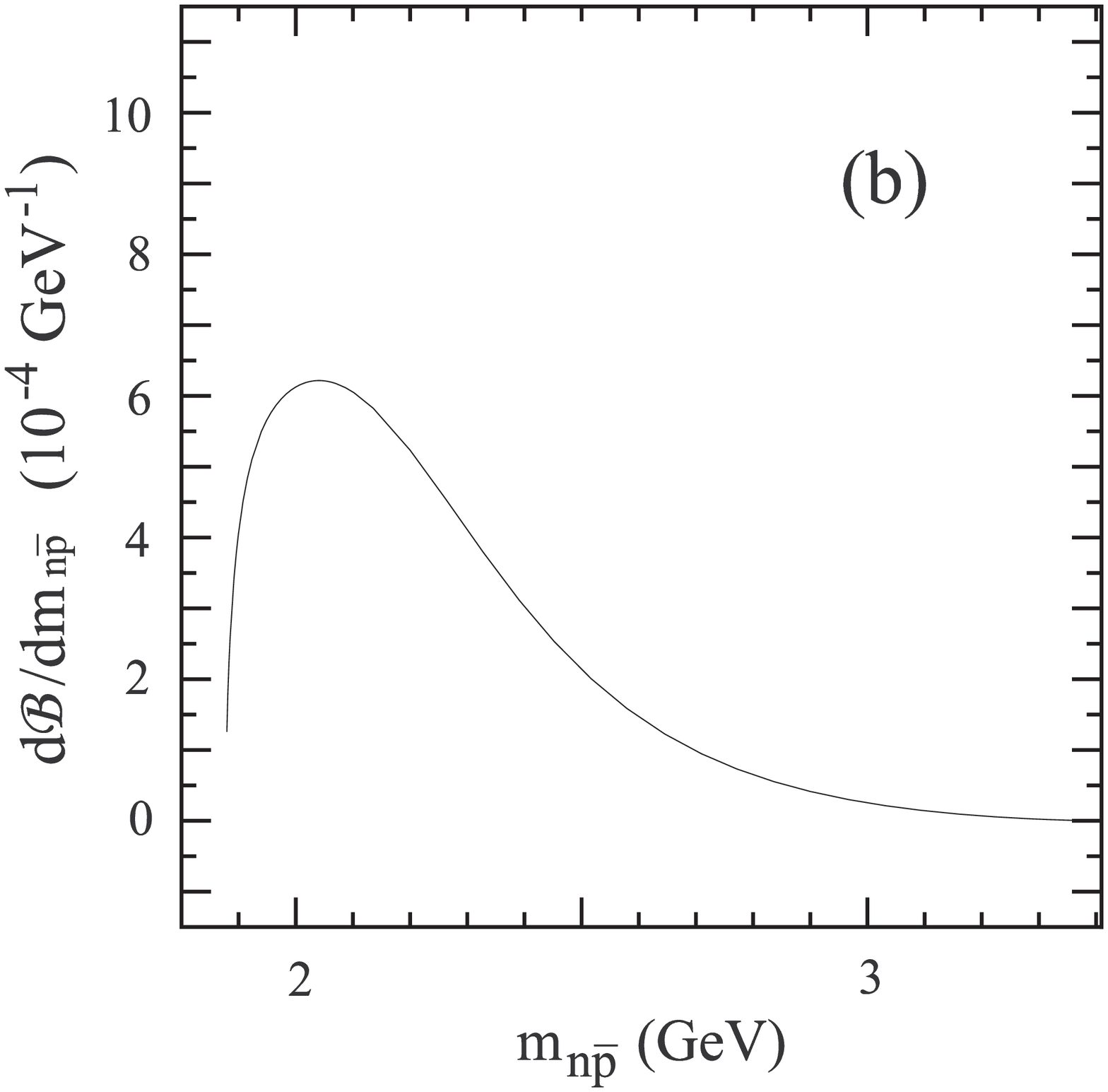,height=2.2in}}   \\[4mm]
    \caption{{\small The $n\bar p$ invariant mass distribution $d\B/dm_{n\bar p}$ of (a) $\ov
B^0\to D^{*+}n\bar p$ and (b) $\ov B^0\to D^+n\bar p$.
    }}
\end{tabular}
\end{figure}

Also shown in Fig. 4 are the $p\bar p$ invariant mass
distributions of $\ov B^0\to D^{*0}p\bar p$ and $\ov B^0\to
D^0p\bar p$ calculated in the MS model. It is clear that the
spectrum of $D^{*0}p\bar p$ is similar to that of $D^{(*)+}n\bar
p$. As for the differential rate of $D^0p\bar p$, it has a hump
around $m_{p\bar p}\sim 3.0$ GeV, which is caused  by the mass
term $q^\mu q^\nu/m_{a_1}^2$ in the propagator of the $a_1$
meson.\footnote{The $\ov B^0\to D^0p\bar p$ spectrum is somewhat
sensitive to the model for form factors. In the NRSX model, the
peak appearing at low $p\bar p$ invariant mass $\sim 2$ GeV is
lower than the hump at $m_{p\bar p}\sim 3$ GeV. This is
inconsistent with experiment [see the data shown Fig. 4(b)].} We
see that the predicted spectrum is consistent with experiment.

\begin{figure}[tb]
\begin{tabular}{cc}
\vspace{0cm}
\hspace{-8cm}\scalebox{1}{\epsfig{file=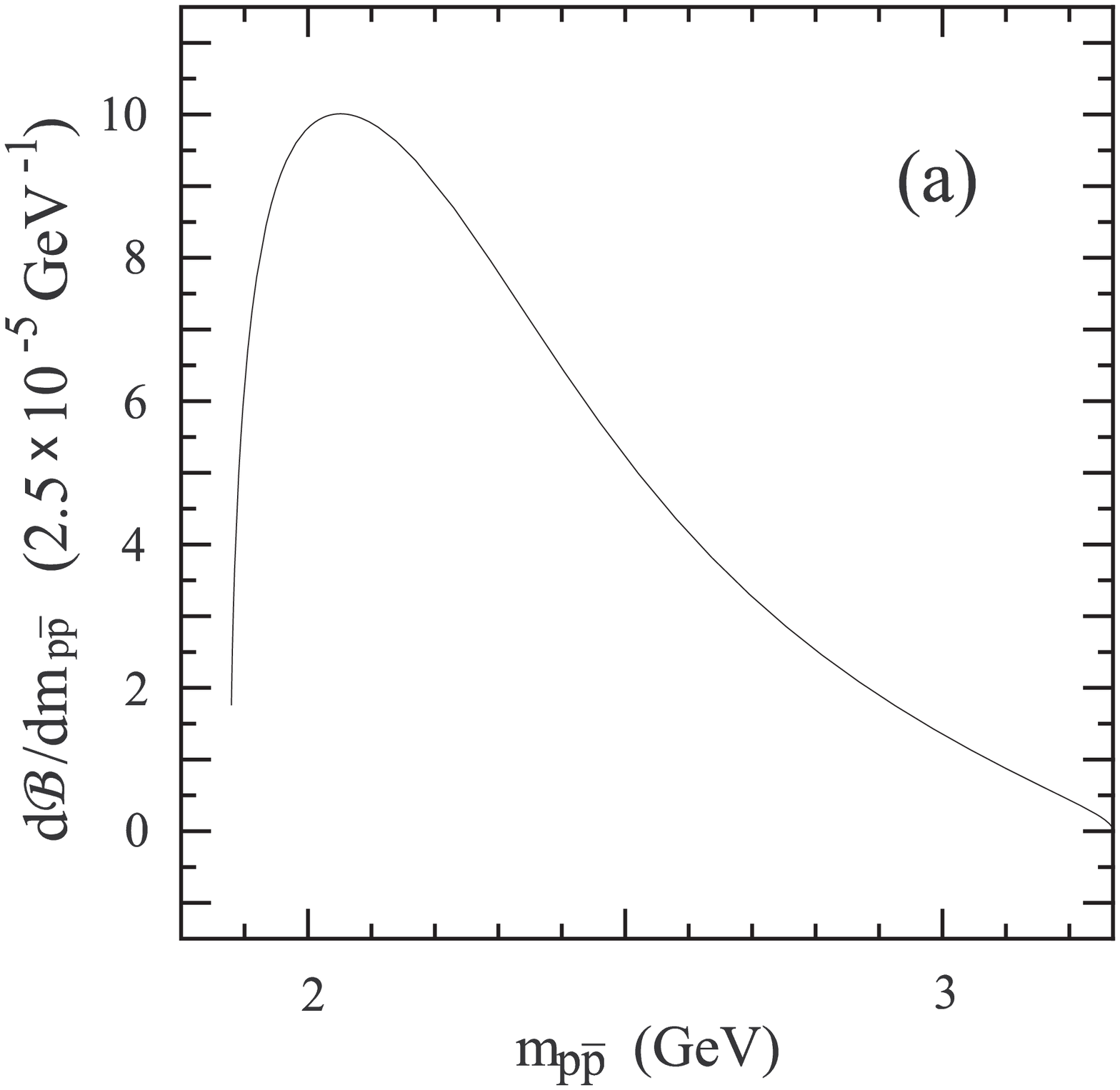,height=2.2in}}
& \hspace{-8cm}
\scalebox{1}{\epsfig{file=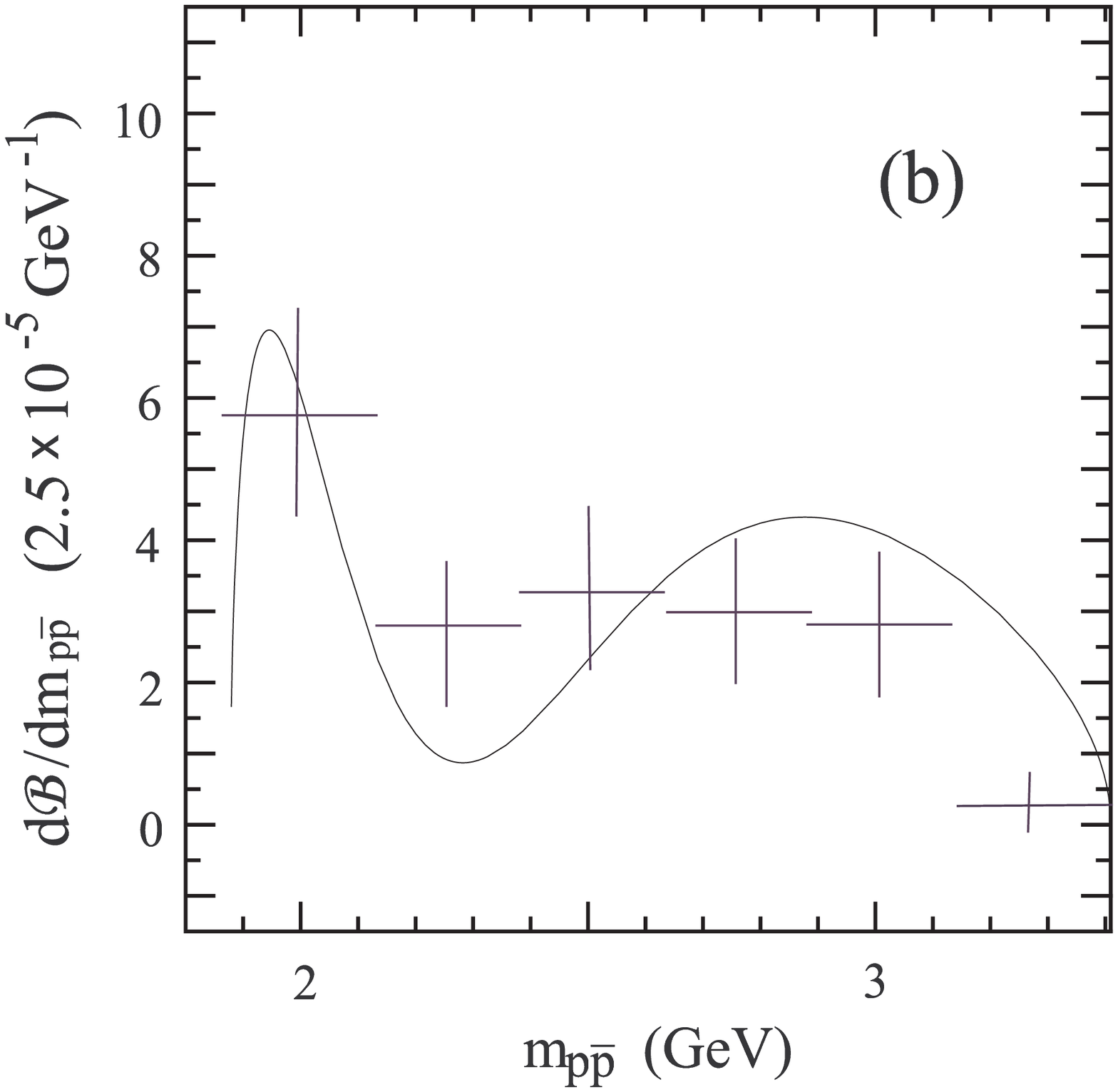,height=2.2in}}   \\[4mm]
    \caption{{\small The $p\bar p$ invariant mass distribution $d\B/dm_{p\bar p}$ of (a) $\ov
B^0\to D^{*0}p\bar p$ and (b) $\ov B^0\to D^0p\bar p$. The
experimental data for the spectrum of $\ov B^0\to D^0p\bar p$ are
taken from [3].
    }}
\end{tabular}
\end{figure}

\section{Conclusions}
We have studied the charmful three-body baryonic $B$ decays $\ov
B\to D^{(*)}N\ov N$: the color-allowed modes $\ov B^0\to
D^{(*)+}n\bar p$ and the color-suppressed ones $\ov B^0\to
D^{(*)0}p\bar p$. While the $D^{*+}/D^+$ production ratio is
predicted to be of order 3, it is found that $D^0p\bar p$ has a
similar rate as $D^{*0}p\bar p$. It is pointed out that $\ov
B^0\to D(D^*)N\ov N$ are dominated by the nucleon vector current
or by vector meson intermediate states, whereas $\ov B^0\to
D^0p\bar p$ proceeds predominately via the axial-vector
intermediate state $a_1(1260)$. The study of the $N\bar N$
invariant mass distribution in general indicates a threshold
baryon pair production, that is, a recoil charmed meson
accompanied by a low mass baryon pair except that the spectrum of
$D^0p\bar p$ has a hump at large $p\bar p$ invariant mass
$m_{p\bar p}\sim 3.0$ GeV. The presence of a hump in the $D^0p\bar
p$ spectrum can be tested by the improved experiment in the
future.

\vskip 3.0cm \acknowledgments  This work was supported in part by
the National Science Council of R.O.C. under Grant Nos.
NSC90-2112-M-001-047 and NSC90-2112-M-033-004.


\end{document}